\documentclass[aps,prc,nofootinbib]{revtex4}
\usepackage{eurosym}
\usepackage{amsmath,amssymb}
\usepackage{graphicx}
\usepackage{dcolumn}
\usepackage{bm}
\usepackage{color}
\usepackage{multirow}
\usepackage{float}
\usepackage{array}
\usepackage{dcolumn}
\usepackage{booktabs}
\usepackage[detect-all]{siunitx}
\usepackage{makecell}
\usepackage{xcolor}
\usepackage{yfonts}
\graphicspath{ {/Users/Nastia/Desktop/research_dis/papers_writing/xenes/nov_5} }

\newcolumntype{P}[1]{>{\centering\arraybackslash}p{#1}}

\begin{document}

\title{%Regularization of the HAL QCD $\Omega N$ Potential and Bound States in $\Omega$-containing Cluster Systems
Bound States  of $\Omega$ Baryons in  Light  Nuclei
}
\author{I. Filikhin$^1$, R. Ya. Kezerashvili$^{2,3,4}$, and B. Vlahovic$^1$}
\affiliation{\mbox{$^{1}$North Carolina Central University, Durham, NC, USA} \\
$^{2}$New York City College of Technology, The City University of New York,
Brooklyn, NY, USA\\
$^{3}$The Graduate School and University Center, The City University of New
York, New York, NY, USA\\
$^{4}$Long Island University, Brooklyn, NY, USA}

\begin{abstract}
\noindent We investigate bound states of light $\Omega_{3x}$-clusters ($x = s, c$),  motivated by 
the $\Omega_{3s}N$ potential recently developed by the HAL QCD collaboration. To regularize this 
potential, we remove the deeply attractive core at $r < 0.4~\mathrm{fm}$ and parametrize 
the long-range component ($r > 0.4~\mathrm{fm}$) using a two-range Gaussian form. 
This procedure preserves the relevant two-body bound state energy while having a negligible 
effect on the $\Omega_{3s}NN$ and $\Omega_{3s}\Omega_{3s}N$ systems.  
An effective $\Omega_{3s}\alpha$ potential is then constructed by fitting a two-range 
Gaussian function to the long-range component of the folding potential, enabling 
calculations of the bound state energies of the  $\Omega_{3s}\alpha$, $\Omega_{3s}\alpha\alpha$ and 
$\Omega_{3s}\Omega_{3s}\alpha$ systems. The regularization procedure leads to a 
substantial reduction in bound state energies compared to those obtained with the 
original potential.  
We further extend the analysis to $\Omega_{3c}$-cluster systems by introducing an 
$\Omega_{3c}N$ interaction, derived by comparing the existing $\Omega_{3s}\Omega_{3s}$ 
and $\Omega_{3c}\Omega_{3c}$ potentials. Our results suggest that several parametrizations 
predict bound states in $\Omega_{3c}$-containing clusters.
Finally, the $\Omega_{3s}\Omega_{3s}$ interaction is described using a contact-like potential approach, motivated by the effective field theory.
\end{abstract}

\maketitle
\date{\today }

\section{Introduction}
The study of exotic hadronic systems, such as dibaryons and tribaryons, remains a central theme in hadron physics. Multiquark states beyond conventional mesons ($q\bar q$) and baryons ($qqq$) have been explored extensively using quark models, lattice QCD, and other theoretical approaches. Historically, the $H$ dibaryon ($uuddss$) was predicted within the MIT bag model~\cite{Jaffe}, and the $\bar{K}NN$ cluster, a meson–baryon–baryon dibaryon, was subsequently proposed~\cite{Kpp}. More complex systems, including three-nucleon clusters with antikaons ($\bar{K}NNN$), have also attracted considerable attention~\cite{Hyodo2012,Gal2016,Kez10,Kez20}.

Although many exotic states have been considered, no quark-level configuration has yet been identified that contains both an $\Omega$ baryon and a nucleon. Nevertheless, strange dibaryons such as $\Omega^- \Omega^-$ and tribaryons like $\Omega^- \Omega^- N$ remain of theoretical interest due to their potential for deep binding. Several studies have suggested the possibility of an $\Omega N$ bound state~\cite{Goldman1987,Li2000,Pang2004,Zhu2015,Huang2015,Etminan2014,Morita2016,LagrangianMethod,Iritani2019,Meissner2017,GV}. While such a system cannot exist as a genuine multiquark configuration, these investigations provide valuable insight into baryon–baryon interactions in the strange sector and impose important constraints on low-energy QCD models.

The formation of tribaryon clusters with strangeness $-3$ and $-6$, such as $\Omega^- NN$ and $\Omega^- \Omega^- N$, remains an open problem. Previous studies~\cite{GV0,GV} employed the Faddeev equations in momentum space with $\Omega N$ potentials derived from lattice QCD~\cite{LagrangianMethod,Iritani2019}, analyzing the $\Omega^- d$ and $\Omega^- NN$ systems. Further calculations using the hyperspherical harmonics method~\cite{Zhang2022,ESE2023} predicted extremely large $\Omega^- NN$ binding energies.

Motivated by these results, the present work investigates light $\Omega^-$ clusters, focusing on the $\Omega^- \alpha$ and $\Omega^- \alpha\alpha$ systems, as well as the mirror system $\Omega^- \Omega^- \alpha$, within the framework of the Faddeev equations in configuration space. Particular attention is given to the short-range behavior of the $\Omega^- N$ potential~\cite{Iritani2019}. We introduce a modification in which the deeply attractive core is cut off at $r_t \approx 0.4$~fm, while the medium- and long-range region ($r > r_t$) is represented by a two-range Gaussian form. This regularization preserves the two-body $\Omega N$ properties but significantly reduces the three-body $\Omega^- NN$ binding energy, thereby mitigating the strong overbinding observed with the original potential. The approach is motivated by the discrepancy between quark-level and hadronic descriptions of these baryonic systems.

To describe the $\Omega^- \alpha$ and $\Omega^- \alpha\alpha$ systems, we construct an effective $\Omega \alpha$ potential by fitting the long-range behavior of a derived $\Omega \alpha$ folding potential with a two-range Gaussian form, based on the modified $\Omega^- N$ interaction. The resulting effective potential is then applied to calculate the ground-state energies of the $\Omega^- \alpha\alpha$ and $\Omega^- \Omega^- \alpha$ systems. A similar approach was used in Ref.~\cite{Etminan2020}, where the long-range behavior was instead fitted with a Woods–Saxon form combined with the original $\Omega^- N$ potential. Since the original potential has Gaussian asymptotics, this mismatch introduced additional uncertainties. Moreover, the strong attraction in the unmodified $\Omega^- N$ interaction tends to overbind the $\Omega^- \alpha\alpha$ and $\Omega^- \Omega^- \alpha$ systems.

Beyond these light strange systems, heavy-quark clusters provide a complementary line of investigation. In particular, the negatively charged, triple-strange $\Omega^-$ baryon (strangeness $-3$, mass $1672.45 \pm 0.29$ MeV/$c^2$) may be compared to its triple-charmed counterpart, the $\Omega_{ccc}^{++}$ baryon (charm $+3$, mass $\sim 5000$ MeV/$c^2$ according to recent lattice QCD calculations~\cite{arXiv:2411.12729v1}). These baryons differ in flavor content, charge, and mass, providing a natural setting for comparative studies. Investigating tribaryon systems such as $\Omega^- \Omega^- N$ and $\Omega_{ccc}^{++} \Omega_{ccc}^{++} N$ thus offers additional insight into the underlying dynamics. Input for such studies is available from HAL QCD calculations of the $\Omega^- \Omega^-$ potential~\cite{Gongyo18} and the $\Omega_{ccc}^{++} \Omega_{ccc}^{++}$ potential~\cite{Yan Lyu}.

In what follows, we adopt the notation $\Omega_{3c}$ for the triple-charmed baryon $\Omega_{ccc}^{++}$ and simply $\Omega$ for the triple-strange baryon $\Omega_{sss}^{-}$. This convention avoids unnecessary complications in notation and reflects our omission of Coulomb effects throughout the discussion.

The paper is organized as follows. Section~\ref{Model} introduces the approach based on Faddeev equations in configuration space and discusses the $\Omega N$, $\Omega \Omega$, $\Omega_{3c}\Omega_{3c}$, $NN$, and $\alpha\alpha$ interactions, including the folding procedure for the $\Omega\alpha$ potential. Section~\ref{sec:5} presents the numerical results and analysis. Concluding remarks are given in Sec.~\ref{sec:6}.

\section{Model}
\label{Model}
\subsection{Approaches}

In our study, the non-relativistic three-body eigenvalue problem is formulated 
using the configuration-space Faddeev equations~\cite{FM} for the components of 
the wave function. We restrict the analysis to three-body systems containing two 
identical particles. The Faddeev decomposition takes the form
\begin{equation}
    \Psi = U + (1\pm P)W,
    \label{FaddeevDec}
\end{equation}
where the components $U$ and $W$ correspond to two possible rearrangements of the 
system: in $U$, the identical particles form a pair interacting with the third 
particle, while in $W$, one of the identical particles interacts with the 
non-identical one. The operator $P$ denotes the permutation of the two identical particles. 
This framework is directly suited to the $\Omega NN$, $\Omega\Omega N$, and 
related cluster systems, where two identical baryons or two $\alpha$ particles 
are present. In the former case, one must choose $-P$ in Eq.~(\ref{FaddeevDec}), 
while in the latter case, the appropriate choice is $+P$.

As an example, the set of the Faddeev equations for three fermions are written as
\begin{equation}
\begin{array}{l}
{(H_{0}+V_{AA}-E)U=-V_{AA}(W-PW)}, \\
{(H_{0}+V_{AC}-E)W=-V_{AC}(U-PW)}.%
\end{array}
\label{GrindEQ__1_}
\end{equation}
Here, $H_{0}$ is kinetic operator of the three-body system, ${V_{AA}}$ and ${V_{AC}}$ represent the interaction potentials between identical fermions
and non-indetical particles.
The spin-isospin variables of the system can be represented by the correspondented basis elements.
After separate of the variables, one can definite 
 the coordinate part, $\Psi^{R}$, of the wave function $\Psi =\xi _{isospin}\otimes \eta _{isospin}\otimes \Psi ^{R}$.

 A well-established numerical approach for solving these equations is 
the finite-difference approximation on a coordinate mesh~\cite{Gignoux1974,FM}. 

In the Faddeev formalism, the particles are treated as structureless, point-like entities. This simplification deviates from the realistic picture, where $\Omega$ baryons and nucleons possess internal quark substructure. While phenomenological nucleon--nucleon potentials are typically adjusted to reproduce scattering data, they are generally unreliable at inter-particle distances shorter than the nucleon radius. Consequently, such two-body potentials often fail to accurately predict three-body observables, such as the binding energies of the $^3$He and $^3$H nuclei. To overcome this limitation, phenomenological three-body forces are frequently introduced to compensate for the breakdown of the quark-cluster picture at short distances. At scales smaller than the nucleon size, the concept of independent particles interacting via pairwise forces becomes invalid. Therefore, it is important to identify the region of inter-particle distances where lattice QCD-based potentials can be reliably applied.  

Another important issue concerns the role of channel coupling in lattice QCD. While most lattice simulations employ a single-channel approximation, multi-channel mixing may occur at short distances due to the underlying quark dynamics. We assume that such mixing becomes significant only at very small separations, where the internal structures of the baryons overlap. This overlap is expected only under extreme conditions, such as high-density nuclear matter or high-energy collisions. Within the Faddeev formalism, however, overlap of baryon wave functions is not permitted.  

Based on this reasoning, we conclude that the direct application of lattice QCD-derived potentials within the Faddeev equations may be unreliable at short distances. Instead, we propose using these potentials only at larger separations---beyond a cutoff radius $r_c$, which we typically choose as the sum of the root-mean-square ($rms$) radii of the interacting particles. The short-range part of the interaction, which is not captured in this approach, can then be reconstructed by smoothly interpolating the potential toward the origin~\cite{GV25}. We define the modified potential as
\begin{equation}
V(r) =
\begin{cases}
V_{\text{interior}}(r), & r < r_c, \\
V_{\text{exterior}}(r), & r \geq r_c,
\end{cases}
\label{V}
\end{equation}
where $V_{\text{exterior}}(r)$ is obtained by fitting lattice QCD data, while $V_{\text{interior}}(r)$ is constructed to smoothly connect with the exterior part. To ensure physical consistency, both the potential and its first derivative are continuous at $r = r_c$, i.e., $V_{\text{interior}}(r_c) = V_{\text{exterior}}(r_c)$ and $V'_{\text{interior}}(r_c) = V'_{\text{exterior}}(r_c)$. This procedure incorporates lattice QCD input while respecting the limitations of quark-level dynamics at short distances. As an example, this reconstruction is applied to the lattice QCD $\Omega N$ potential, which is known to be purely attractive due to the absence of the Pauli exclusion principle between non-identical baryons.  

A similar methodology is applied to $\alpha$-cluster systems, where the effective $\Omega\alpha$ interaction is approximated by a folding potential derived from the $\Omega N$ potential and is valid at distances larger than the radius of the $\alpha$ particle. Systems consisting of $Y$ hyperons and $C$ $\alpha$-clusters, of the form 
$Y\Omega + C\alpha$ with $Y=1,2$ and $C=2,1$, are considered. This approach was previously employed in Ref.~\cite{Etminan2020} to determine the $\Omega\alpha$ effective potential.  

The Faddeev equations take their simplest form in the $s$-wave approximation~\cite{FKV25}, which is adopted in the present work. However, $\alpha$-cluster systems such as $\Omega\alpha\alpha$ cannot be described within an $s$-wave model alone. This limitation arises from the Pauli principle, which suppresses the $\alpha\alpha$ $s$-wave component and requires the inclusion of higher partial waves, most notably the $d$- and $g$-waves~\cite{FSV05}. In addition, the Coulomb interaction --essential for binding two $\alpha$ particles into the resonant ground state of $^8$Be -- must also be included when treating $2\alpha$-cluster systems.

In the case of the $\Omega\Omega$ interaction, strong short-range repulsion appears in addition to attraction at intermediate distances. The repulsive core is a manifestation of the Pauli exclusion principle acting between identical fermions. This interaction can be represented as the sum of repulsive and attractive contributions:
\begin{equation}
V(r) = V_{\text{rep}}(r) + V_{\text{attr}}(r),
\label{VV}
\end{equation}
where $V_{\text{rep}}(r)$ describes the repulsive part of the interaction, and $V_{\text{attr}}(r)$ represents the attractive component. It should be emphasized that $V_{\text{rep}}(r)$ is distinct from the short-distance term $V_{\text{interior}}(r)$ introduced in Eq.~(\ref{V}).  

Our three-body calculations presented below employ the $\Omega\Omega$ potential obtained directly from lattice QCD. We further revise this potential to isolate the role of the repulsive core, separating it into repulsive and attractive parts as in Eq.~(\ref{VV}). The repulsive component is modeled as a contact-like interaction, motivated by effective field theory~\cite{W}, to capture short-distance physics.

\subsection{$\protect\Omega  N$, $\Omega  \Omega $,  and $\Omega_{3c}  \Omega_{3c} $ potentials}
\label{interraction}
%M. Yamada, et al., (HAL QCD Collaboration), Prog. Theor. Exp. Phys. 2015, 071B01 (2015).

To seek for a possible dibaryon states in the strangeness $-3$ channel in
Ref. \cite{Etminan2014} authors calculated the $\Omega N$ potential through
the equal-time Nambu--Bethe--Salpeter wave function in (2 +1)-flavor lattice
QCD with the renormalization group. By solving the Schr\"{o}dinger equation
with this potential, authors found one bound state with binding energy 18.9
MeV in state $^{5}S_{2}$. Recently, in Ref. \cite{Iritani2019} $\Omega N$ in
the $S-$wave and spin-2 channel is studied from the (2+1)-flavor lattice QCD
with nearly physical quark masses ($m_{\pi }=$ 146 MeV and $m_{K}=$ 525MeV)
by employing the HAL QCD method. The $\Omega N$ ($^{5}S_{2}$) potential,
obtained under the assumption that its couplings to the $D-$wave
octet-baryon pairs are small, is found to be attractive in all distances and
produces a quasi-bound state 1.54 MeV for $n\Omega ^{-}(uddsss)$ and 2.46
MeV for $p\Omega ^{-}(uudsss)$. In the later case the binding energy
increase is due to the extra Coulomb attraction. The fitted lattice QCD
potential by Gaussian and Yukawa squared form for obtained observables such
as the scattering phase shifts, root mean square distance, and binding
energy, has the form \cite{Iritani2019}:
\begin{equation}
V_{\Omega N}=b_{1}e^{-b_{2}r^{2}}+b_{3}\left( 1-e^{-b_{4}r^{2}}\right)
\left( \frac{e^{-m_{\pi }r}}{r}\right) ^{2}.  \label{HALInteraction}
\end{equation}%
 Four sets of the
fitting parameters $b_{1}$, $b_{2}$, $b_{3}$ and $b_{4}$ are found from the
simulation \cite{Iritani2019} and the pion mass is $m_{\pi }=146$ MeV. The parameter sets $Pi$, $i$=1,2,3,4 are presented in Ref. \cite{Iritani2019}.
Following \cite{GV}, we use the P$_1$ set of parameters: $b_{1}$=-306.5 MeV, $b_{2}$=73.9 fm$^{-2}$, $b_{3}$=-266 MeV fm$^2$ and $b_{4}$=0.78 fm$^{-2}$. 
%\begin{table}[t]
%\caption{\label{tab:2} The parameters of  $P_i$, $i$-1,2,3,4 potentials for  the
%$^5S_2$ $\Omega N$ interaction.}
%\begin{tabular}{lcccc}  \hline\noalign{\smallskip}  
% & $P_1$ & $P_2$ & $P_3$ & $P_4$ \\
% \hline\noalign{\smallskip}  
%$b_1$ (MeV)            &  -306.5  &  -313.0  &  -316.7  & -296  \\
%$b_2$ (fm$^{-2}$)      &      73.9  &      81.7  &      81.9  &     64  \\
%$b_3$ (MeV fm$^{2}$)   &  -266    & -252    &  -237    & -272  \\
%$b_4$ (fm$^{-2}$)      &       0.78 &       0.85 &       0.91 &      0.76 \\  \hline\noalign{\smallskip}  
%\end{tabular}
%\end{table}
The scattering characteristics obtained with the sets  are
found to be consistent with each other within statistical errors  \cite{GV}. The Yukawa
squared form at long distance is motivated by the two-pion exchange between $%
N$ and $\Omega $.
It was reported in 1987 \cite{Goldman1987} that $%
\Omega N $ is\ the most promising candidate of a stable dibaryon and it is a
possibility to have a bound state in the $S$-wave and total-spin 2 channel 
\cite{Etminan2014}.

The $\Omega\Omega$ is the most exotic dibaryon system with strangeness $-6$. So far, only the deuteron is known as a stable dibaryon, formed by a proton and neutron in the spin-triplet, isospin-singlet channel. The $\Omega\Omega$ structure was first studied in the chiral SU(3) quark model using the resonating group method \cite{Zhang2000}, which predicted a bound state and proposed searches in heavy-ion collisions. Lattice QCD simulations with $m_\pi=146$ MeV, $a\simeq0.0846$ fm, and volume $8.1$ fm$^3$ later reported the first results and physical implications \cite{Gongyo18}.

The Pauli principle allows only the $^{1}S_{0}$ and $^{5}S_{2}$ $S$-wave states for the $\Omega_{3c}\Omega_{3c}$ system. 
HAL QCD results~\cite{QCD2015} indicate a strong attraction in the $^{1}S_{0}$ channel. 
Studies of the most strange dibaryon, $\Omega\Omega$, also indicate strong attraction in the $^{1}S_{0}$ channel~\cite{Gongyo18}. 
Both dibaryons have been investigated within $(2+1)$-flavor lattice QCD with nearly physical quark masses in Refs. \cite{Gongyo18} and \cite{Yan Lyu}. 
The corresponding potentials are parametrized as a sum of three-range Gaussians:
\begin{equation}
V_{\Omega_{3x}\Omega_{3x}}(r) = \sum_{i=1}^{3} \alpha_{i}\,\exp\!\left[-\beta_{i}r^{2}\right],
\label{OmegaOmega}
\end{equation}
where $x=s,c$ and  fitting parameters listed in Table~\ref{ParOmegas}. 
The central potential~(\ref{OmegaOmega}) reproduces scattering phase shifts and binding energies, and it features short-range repulsion together with intermediate-range attraction, but no long-range pion-exchange tail.
\begin{table}[h!]  
\caption{\label{tab:Fit_Omega3c} Fitting parameters of  $ \Omega_{3c}\Omega_{3c}$ and $\Omega_{3s}\Omega_{3s}$ potentials both in spin channel $^1S_0$  given in Eq.~\eqref{OmegaOmega} at lattice Euclidean time $ t/a = 26$ and $t/a$=17, respectively.   These parameters are taken from Ref.~\cite{Yan Lyu} and adapted from Ref.~\cite{Gongyo18}.
}
\centering
\begin{tabular}{ccccccccc}
\hline
\noalign{\smallskip}
&$x$& $\alpha _{1},$ MeV & $\beta _{1}, $ fm & $\alpha _{2}$ & $\beta _{2}, $ fm & 
$\alpha _{3}$, MeV & $\beta _{3}, $ fm 
\\ \hline
\noalign{\smallskip}
\cite{Yan Lyu}&$\Omega _{3c}\Omega _{3c}$ & 239 & 48.5 & $-62.7$ & 7.8 & $-98.8$ & 3.4   \\ \hline
\noalign{\smallskip}
\cite{Gongyo18}&$\Omega_{3s}\Omega_{3s}$ & 914 & 48.9 & 305 & 10.8 & $-112$ & 1.1   \\ \hline
\noalign{\smallskip}
\end{tabular}
\label{ParOmegas}
\end{table}

\subsection{$NN$ and $\alpha \alpha$ interactions}
For description of the nuclon-nucleon interaction, we use the phenomenological MT-I-III \cite{Malfliet1969} potentials  with the paremeters corrected in Ref. \cite{Friar}.

The interaction between two $\alpha$ particles is modeled as the sum of nuclear and Coulomb components:
\begin{equation}
V_{\alpha\alpha}(r) = V_n(r) + V_C(r).
\label{aa}
\end{equation}
The nuclear part, $V_n(r)$, is described by the phenomenological local Ali–Bodmer potential~\cite{AliBodmer}, commonly used to reproduce $\alpha\alpha$ scattering data. In this work, we adopt the parameter set, which defines the nuclear potential as a superposition of one-, two- and three-range Gaussians.
This potential acts in partial waves with orbital angular momentum quantum numbers $l = 0, 2, 4$, as appropriate for the $\alpha \alpha$ system. Parametrizations of the  $\alpha \alpha$  potential of the Ali-Bodmer type are presented in Table \ref{tab:Vn}.
\begin{table}[h!]
\centering

\caption{Parametrizations of the  $\alpha \alpha$  potential $V_n(r)$=$V^{rep}_n(r)+V^{attr}_n(r) $ for different orbital angular momenta $l$. The Ali-Bodmer potential sets  from \cite{AliBodmer}. %are presented. 
}
\begin{tabular}{l|ll}
 \hline
 \noalign{\smallskip}  
 \centering
$l$ & $V^{rep}_n(r)$& $V^{attr}_n(r) $ \\
 \hline\noalign{\smallskip}  
0 (set a) & $125.0 e^{-(r/1.53)^2}$&$ - 30.18 \, e^{-(r/2.85)^2}$  \\
2 (set a) & $20.0 e^{-(r/1.53)^2}$&$ - 30.18 \, e^{-(r/2.85)^2}$ \\
4 (set d) &0 &$-130.0 \, e^{-(0.475 r)^2}$ \\
\hline
\end{tabular}
\label{tab:Vn}
\end{table}
The Coulomb potential $ V_C(r)$ is only used  in calulations related to the $\alpha \alpha$ interaction.

\subsection{Folding $\Omega \alpha$ potential}

The effective $\Omega \alpha $ interaction is obtained using a
single-folding potential method \cite{Satchler1983,Miyamoto2018}.  Recently, in Ref.  \cite{Filikhin2025Fold}, such potential has
been  constructed by using a $\Omega N$ HAL QCD potential in state $^{5}S_{2}$ \cite{Iritani2019} as an effective interaction. 
The  folding $\Omega \alpha$ potential, $V_{\Omega\alpha}(r)$, is calculated using the formula:
$$
V^f_{\Omega \alpha}(r)=\int_{-1}^1du\int_0^{R_{max}}dx \rho(x)V_{\Omega N}(\sqrt{x^2+r^2-2xru})x^2
$$
where $\rho(x)$ denotes the normalized density distribution of nucleon matter in the $\alpha$ particle.

The folding potential can be fitted using the Woods--Saxon (WS) form, which is assumed 
to describe the asymptotic region $r > 2.0$~fm, i.e., beyond the known root-mean-square 
($rms$) radius of the $\alpha$ particle~\cite{Etminan2020}. Unlike Ref.~\cite{Etminan2020}, 
we employ a two-range Gaussian function to fit the folding potential for $r \geq 2.0$~fm. 
This choice is motivated by the asymptotic behavior of the $\Omega N$ potential, which 
itself has a Gaussian form, and by our model expressed in Eq.~(\ref{V}), where $r_c$ 
is related to the $rms$ radius of the $\alpha$ particle.  

As discussed in Ref.~\cite{Wang2024}, the density function $\rho(r)$ can be represented 
by a simple Gaussian matter distribution,  
\[
\rho(r) = \left( \frac{A^{2}}{\pi }\right) ^{3/2} e^{-C^{2}r^{2}},
\]  
which yields  
\[
\langle r^{2}\rangle^{1/2} = \sqrt{\tfrac{3}{2C^{2}}},
\]  
and reproduces experimental data using the parameters given in Ref.~\cite{Wang2024}.  
The $rms$ radius of the $\alpha$ particle is taken to be 1.7~fm, in agreement with 
near-threshold $\phi$-meson photoproduction data from the LEPS Collaboration 
\cite{HiraiwaLEPS}, which report an $rms$ matter radius of ${}^4$He equal to 
$1.70 \pm 0.14$~fm~\cite{Wang2024}.

\section{Numerical Results}
\label{sec:5}
\subsection{Fitting for HAL QCD   $\Omega  N$ potential }
\begin{figure}[ht]
\begin{center}
\includegraphics[width=25pc]{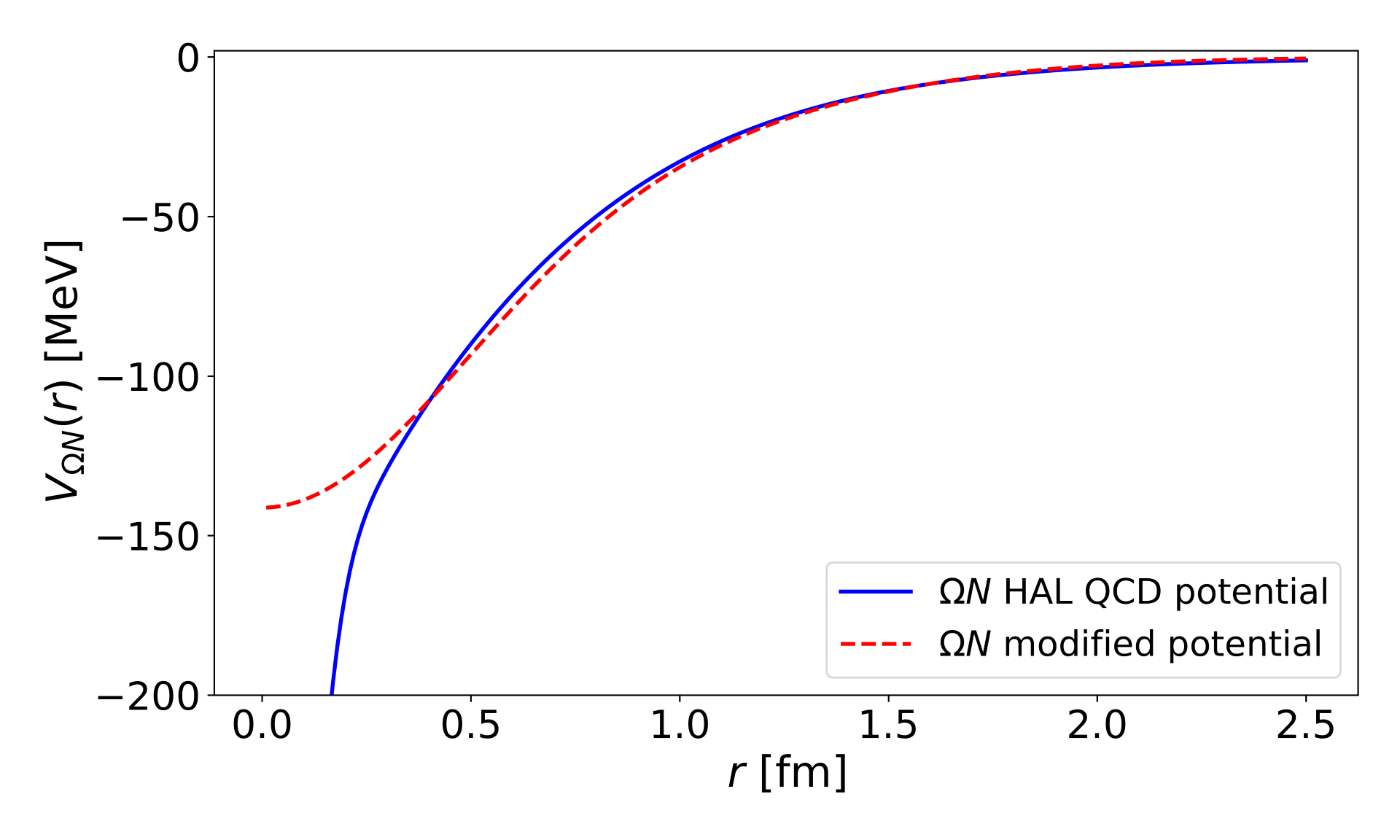}
\end{center}
\caption{$\Omega N$ potentials. The solid curve represents the HAL QCD $\Omega N$ potential ${V}_{\Omega N}$ with parameters P$_1$~\cite{Iritani2019}, while the dashed curve shows the renormalized potential $\widetilde{V}_{\Omega N}$ given by Eq.~(\ref{P1t}).
}
\label{fig:1}
\end{figure}

Investigations of the $\Omega$-clusters within a nonrelativistic potential model require corresponding interactions potentials. In particular, %cluster 
$\Omega\alpha$ interaction is based on a $\Omega N$ potential. According to our method, we propose a modification for the HAL QCD potential to take into account the particle-frame of interactions in the ligth $\Omega$-clusters.
This means that we renormalized the $\Omega N$ potential to remove the strong 
attraction near the origin. The asymptotic region is taken to begin at $r = 1.2$~fm, 
which is defined as the sum of the root-mean-square radii of the nucleon and the 
$\Omega$ baryon.
In this region, the  $\Omega N$ potential is fitted by Gaussian functions, as shown in Fig.~\ref{fig:1}. 
The resulting potential eliminates the excessively strong attraction attributed to the original HAL QCD potential.
%The effect of such renormalization can be evaluated by the data listed in Table \ref{t22}. One can see that the reving of the exteremal atraction doe not change the two-body energy $E_2$ and scattering parameters.

The effect of such renormalization can be evaluated from the data listed in Table~\ref{t22}. 
It can be seen that removing the extreme short-range attraction does not change the two-body energy $E_{2}$ or the scattering parameters.
\begin{table}[!ht]
\caption{
Low-energy characteristics of the $\Omega N$ interactions.  $E_2$ is the ground state energy,
 $a_{\Omega N}$ is the scattering length, and  $r_{\Omega N}$ is the effective range parameter. Calculations performed with HAL QCD $V_{\Omega N}$ with parameters P1 and potential (\ref{P1t}). 
}
\label{t22}
\begin{tabular}{cccc} \hline\noalign{\smallskip}
Potential & $a_{\Omega N}$ (fm) & $r_{\Omega N} $ (fm)& $E_2$ (MeV) \\
\hline\noalign{\smallskip}
 $V_{\Omega N}$  & 6.4  & 1.3 & -1.29 \\
  $V_{\Omega N}$  & -- & -- & -1.29 \cite{GV}   \\
% $\Omega N$, P1 &  -- & -- & -1.54 \cite{Etminan2020}  \\ 
%$\Omega N $, 
$\widetilde{V}_{\Omega N}$  & 6.2 & 1.2 & -1.29 \\
% $\Omega_{3c}N $,  P1t & 2.8  & 1.0 & \\
\noalign{\smallskip}\hline
\end{tabular}
\end{table}
The parametrization of the $\widetilde{V}_{\Omega N}$ potential is given by the expression:  
\begin{equation}
\widetilde{V}_{\Omega N}(r) = -80.289\, e^{-(r / 0.634)^2} 
- 60.959 \, e^{-(r / 1.131)^2}.
\label{P1t}
\end{equation}
\begin{table}[!ht]
\caption{Two- and three-body ground state energies in the system $\Omega NN$ and $\Omega\Omega N$. $E_3$ are ground state energies of the $\Omega NN$ system ($s=5/2$, $t=0$) and $\Omega\Omega N$ system ($s=1/2$), respectively, including the cases where $m_\Omega = m_{\Omega_{3c}}$. The parameter $\gamma$ is a scaling factor of the HAL QCD $V_{\Omega N}$ potential with parameters P$_1$. 
$\widetilde{V}_{\Omega N}$ is a two-range Gaussian (Eq.~(\ref{P1t})) simulating the P$_1$ parametrization of $V_{\Omega N}$ in the range 0.6–3.0~fm of the inter-particle distance. Energies are in MeV.
%$\widetilde{V}_{\Omega N}$ is a two-range Gaussian potential (\ref{P1t}) simulating HAL QCD $V_{\Omega N}$ with P1 parameters in the range 0.6--3.0~fm of the inter-particle distance. The energies are given in MeV.
}
\label{t111}
\begin{tabular}{lcccc} \hline\noalign{\smallskip}
System &Potential & $E_2^{\Omega N}$ &$E_2^{NN}$ & $E_3$  \\
\hline\noalign{\smallskip}
$\Omega NN$ &$V_{\Omega N}$ & -1.29&-2.23 &  -19.58   \\     
 & $V_{\Omega N}$ & -1.29&-2.23 & -19.6 \cite{GV}  \\    
&$\widetilde{V}_{\Omega N}$& -1.29&-2.23 & -19.92   \\                            
 $\Omega NN$, $m_\Omega=m_{\Omega_{3c}}$&$\widetilde{V}_{\Omega N}$& -5.81&-2.23  &-32.91\\    \hline\noalign{\smallskip}
% & & -21 \cite{Etminan2020}\\  
System &Potential & $E_2^{\Omega N}$ &  $E_2^{\Omega\Omega}$ & $E_3$  \\
\hline
\noalign{\smallskip}
$\Omega\Omega N$ &$V_{\Omega N}$& -1.29& -1.41&  -6.00    \\
                                &    $V_{\Omega N}$             & -1.29  &-1.6&  -6.0\cite{GV}\\
                                & $\widetilde{V}_{\Omega N}$& -1.29& -1.41&  -6.15\\          
$\Omega\Omega N$, $m_\Omega=m_{\Omega_{3c}}$& $V_{\Omega N}$ &-5.813 &-5.539   &-23.46    \\ 
%                                                  &          P1   &-- &-5.68\cite{}  &--  \\           
                                                    &      $\gamma V_{\Omega N}$, $\gamma$=0.8  &-1.463&-5.539  &-10.04     \\                                                                        
                                                   &       $\gamma V_{\Omega N}$, $\gamma$=0.7  &-0.3157&-5.539  &-5.78\\                                                                                                 
&$\widetilde{V}_{\Omega N}$&-5.850 &-5.539   &-23.45   \\    
 \noalign{\smallskip}\hline
\end{tabular}
\end{table}
 The results show only minor differences between the binding energies obtained with the $V_{\Omega N}$ and $\widetilde{V}_{\Omega N}$ potentials, namely 0.3~MeV for $\Omega NN$ and 0.15~MeV for $\Omega\Omega N$.
In other words, the strong attraction near the origin of the HAL QCD $\Omega N$ potential has no significant effect on the ground-state binding energy of the systems.
Replacing the $\Omega$ mass with the $\Omega_{3c}$ mass (about three times larger) 
increases the binding energies by factors of roughly $3/2$ and 4 for $\Omega NN$ and $\Omega\Omega N$ system, respectively.
The dependence on the scaling factor $\gamma$ ($\gamma {V}_{\Omega N}$) further indicates that the 
$\Omega\Omega N$ bound state energy $E_3$ is quate sensitive to the baryon–nucleon 
interaction strength: a 30\% reduction lowers $E_3$ to the two-body $E_2^{\Omega\Omega}$ 
threshold. Finally, we note that our results for the miror  systems $\Omega NN$ and $\Omega\Omega N$ agree well with those of Ref.~\cite{GV}, 
obtained in momentum space formalism. Variations of the $\Omega N$ potential, denoted as P$_1$, P$_2$, P$_3$, and P$_4$ in Ref \cite{GV0,GV}, yield $\Omega\Omega N$ binding energies of 6.00, 6.31, 5.86, and 6.22 MeV, respectively. These values are comparable to the corresponding results reported in Ref.~\cite{GV}: 6.0, 6.2, 5.9, and 6.1 MeV.

\subsection{ $\Omega_{3c} N$ potential modeling}

In this section, we model the $\Omega_{3c}N$ potential by analogy with the $\Omega N$ interaction, under the assumption that similar parallels hold between the $\Omega\Omega$ \cite{Gongyo18} and $\Omega_{3c}\Omega_{3c}$ \cite{Yan Lyu} potentials. 
\begin{table}[!ht]
\caption{
Low-energy characteristics of the $YY$ interactions, where $Y=\Omega, \Omega_{3c}, \Lambda, \Xi^0 $.  
}
\label{t22a}
\begin{tabular}{lccc} \hline\noalign{\smallskip}
Potential & $a_{YY}$ (fm) & $r_{YY} $ (fm)& $E_2$ (MeV) \\
\hline\noalign{\smallskip}
 $\Omega \Omega $   & 6.4  & 1.3 & -1.41 \\
 $\Omega \Omega $ \cite{Gongyo18}  & 4.6(6)   & 1.27(3) & -1.6(6) \\
 $\Omega_{3c} \Omega_{3c}$\cite{Yan Lyu}   & 1.57 & 0.57 &  -5.68 \\ \hline \noalign{\smallskip} 
 $\Lambda\Lambda$ \cite{FSV09}(e)& -0.5  &10.4& UNB\\
$\Xi  \Xi$ \cite{GVV16}& -7.2 & 2.0 & UNB\\
\noalign{\smallskip}\hline
\end{tabular}
\end{table}
%{\color{blue}{
%It is important that the potential reproduce physical observables such as the scattering phase shifts, scattering length, and binding energy. 
The potential must reproduce physical observables such as the scattering phase shifts, scattering length, and binding energy. In Table~\ref{t22a}, the low-energy characteristics of $\Omega$-baryon interactions, namely the scattering length, effective range, and binding energy, are presented. The binding energy of the $\Omega\Omega$ dibaryon obtained with potential~(\ref{OmegaOmega}) is $1.54$~MeV. For comparison, the two-body results for the $\Lambda$ and $\Xi^0$ ($\Xi$) baryons are also shown. The $\Lambda\Lambda$ and $\Xi\Xi$ systems are unbound and have negative scattering lengths. Moreover, the $\Omega\Omega$ dibaryon exhibits the smallest effective range. Thus, the $\Omega\Omega$ attraction is strong enough to bind a pair of $\Omega$ baryons.
\begin{figure}[ht]
\begin{center}
\includegraphics[width=25pc]{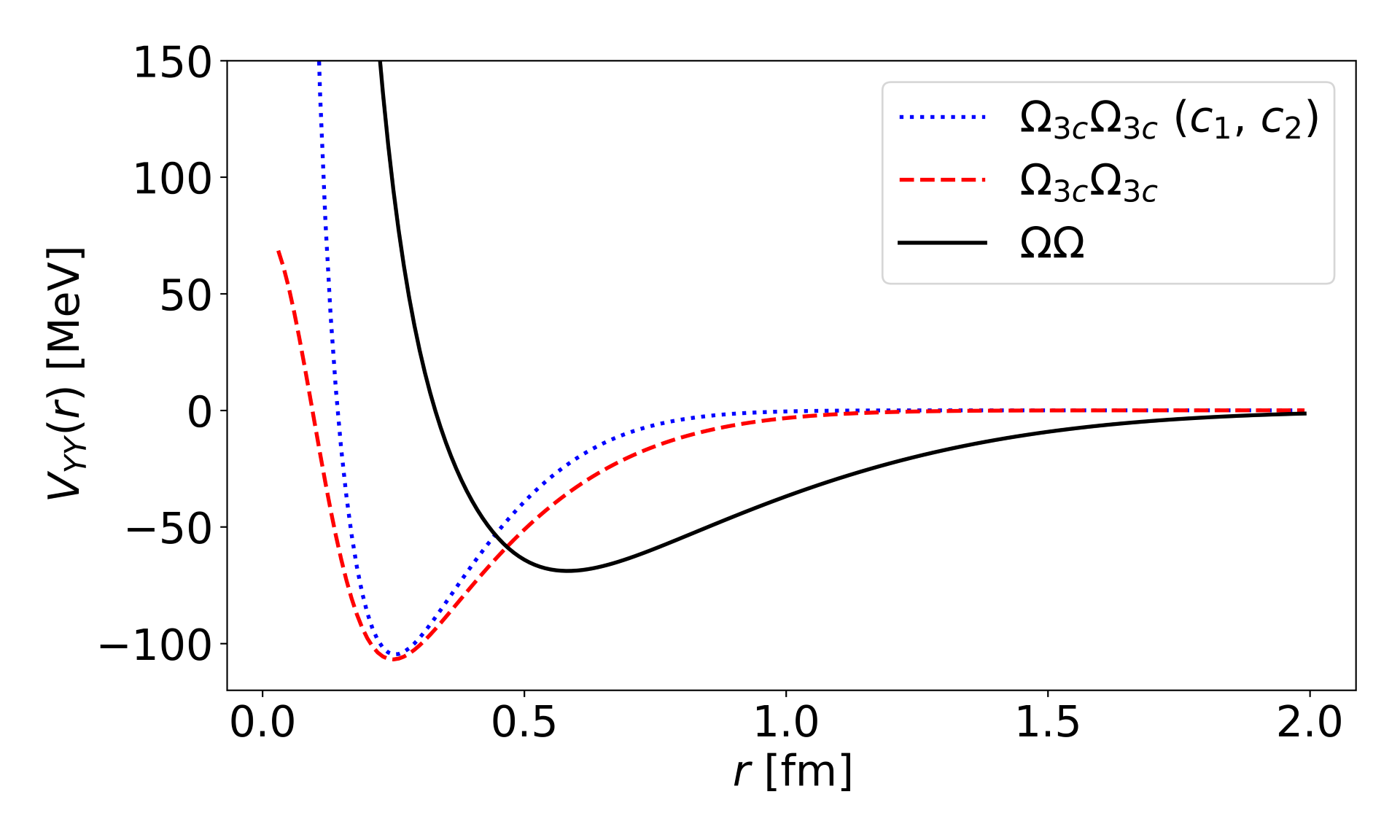}
\end{center}
\caption{$\Omega\Omega$ (solid curve) and $\Omega_{3c}\Omega_{3c}$ (dashed curve) potentials $V_{YY}$ ($Y$ means $\Omega$ or $\Omega_{3c}$). The approximation of Eq. (\ref{Vccc}) for $\Omega_{3c}\Omega_{3c}$ potential is show for $c_1$=2.3 and $c_2$=1.52 by the doted curve. 
}
\label{fig:1a}
\end{figure}

As mentioned in \cite{Yan Lyu}, the $\Omega_{3c}\Omega_{3c}$ interaction has the same qualitative
behaviors as the $\Omega\Omega$ potential \cite{Gongyo18}. Let us consider 
the relation between the $\Omega\Omega$ and $\Omega_{3s}\Omega_{3}$ potentials as 
\begin{equation}
V_{\Omega_{3c}\Omega_{3c}}(r)=c_2V_{\Omega\Omega}(c_1r).
\label{VOccc}
\end{equation}
The relation (\ref{VOccc}), though approximate, provides a basis for 
constructing an $\Omega_{3c}N$ potential by analogy with the $\Omega N$ 
interaction. We adopt a modified form of the $\Omega N$ potential, 
$\widetilde{V}_{\Omega_{3c}N}$, introducing also two parameters: 
\begin{equation}
\widetilde{V}_{\Omega_{3c}N}(r) = c_2 \widetilde{V}_{\Omega N}(c_1 r).
\label{Vccc}
\end{equation}
In (\ref{VOccc}) and (\ref{Vccc}),  parameters $c_1$ and $c_2$ are free parameters. The parameter $c_1$ accounts the radial shift relative to the $\Omega\Omega$ or $\Omega N$ cases, and $c_2$ controls the interaction strength.
Fig. \ref{fig:1a} presents the comparison of the interaction $\Omega_{3c}\Omega_{3c}$($c_1,c_2$)  (\ref{VOccc})  with HAL QCD $\Omega_{3c}\Omega_{3c}$ \cite{Yan Lyu} and $\Omega \Omega$ \cite{Gongyo18} potentials. 

\begin{table}[!ht]
\caption{Ground-state energies $E_3$ of the $\Omega_{3c}\Omega_{3c}N$ and $\Omega_{3c}NN$ systems with $\Omega_{3c}N$ potentials from Eq.~(\ref{Vccc}); parentheses denote results without identical-particle interactions, $E_2^{YN}$ gives the two-body energy with corresponding scattering parameters, and $\delta$ from Eq.~(\ref{d}) represents mass polarization. UNB means ``unbound''.}
\label{t:22}
\begin{tabular}{ccccccccc} \hline\noalign{\smallskip}
System &$c_1$ & $c_2$  & $a_{YN}$ (fm) & $r_{YN} $ (fm)&$E_2^{YN}$ (MeV) &$E_3$ (MeV)& $\delta$ \\
\hline\noalign{\smallskip}
$\Omega_{3c} NN $  &1.35& 1.3& 7.61& 1.00& -0.7638 & -18.26 (-1.597)& 0.04\\
                               &1.35& 1.2&25.4 & 1.07& -0.1163 & -14.51 (-0.284)& 0.18\\   
                                                              &1.41& 0.52&-0.57 & 1.77& UNB & UNB& --\\ 
                               \hline\noalign{\smallskip}                                                              
$\Omega_{3c} \Omega_{3c} N $ &1.1& 1.3&  2.29& 0.89 & -9.740&-35.35 (-32.42)&0.35 \\
                                &1.2& 1.3& 2.99& 0.95& -4.926 & -23.28 (-20.50)&0.51\\
                                &1.3& 1.2& 8.15& 1.05& -0.661 & -9.129 (-6.640)&0.80\\                                                                                                                                  
                                &1.35& 1.3& 7.61& 1.00& -0.7638 &-5.782  (-7.401)&0.79\\ 
                                &1.35& 1.2&25.4 & 1.07& -0.1163 & -6.382 (-3.750)&0.95\\   
                                 &1.41& 0.52&-0.57 & 1.77& UNB & UNB& --\\ 
\noalign{\smallskip}\hline
\end{tabular}
\end{table}
In Table~\ref{t:22}, we show results obtained by varying the parameters 
\(c_1\) and \(c_2\) in the modified \( \Omega N \) potential~(\ref{Vccc}), 
specifically for systems involving the \( \Omega_{3c} \) baryon and nucleons. 
We report the scattering length \( a_{\Omega_{3c}N} \) for the interaction 
between the \( \Omega_{3c} \) baryon and the nucleon. 
The scattering length characterizes the low-energy behavior of the interaction; 
a large positive value typically signals a near-threshold or weakly bound state. 
The effective range \( r_{\Omega_{3c}N} \) provides additional information 
on the shape of the potential beyond its core.  

For the \( \Omega_{3c}NN \) system, a bound state of a proton and a neutron 
(\(N\!N = np\)) is assumed. 
In Table \ref{t:22}, values in parentheses correspond to results obtained when the interaction 
between identical particles is neglected. These serve as estimates of the 
relative strength of the attraction associated with the corresponding potentials. 
We observe that the $NN$ potential exhibits stronger attraction than the 
$\Omega_{3c}\Omega_{3c}$ potential, as reflected in the difference 
$E_3(V_{YY}=0) - E_3$, where $Y$ denotes either $N$ or $\Omega_{3c}$.  

Furthermore, we demonstrate the mass-polarization effect~\cite{H2002,FilKez2018} 
for both systems. In particular, the ratio  
\begin{equation}
\delta = \frac{E_3(V_{YY}=0) - 2E^{YN}_2}{E_3(V_{YY}=0)} ,
\label{d}
\end{equation}
is small for the $\Omega_{3c}NN$ system but large for the 
$\Omega_{3c}\Omega_{3c}N$ system. 
In the former case, $E_3(V_{NN}=0) \approx 2E^{\Omega_{3c}N}_2$, 
whereas in the latter case the mass-polarization effect cannot be neglected.  

To sum up, Table~\ref{t:22} illustrates how variations in the \( \Omega_{3c}N \) potential 
influence the low-energy characteristics and binding energies of few-body 
systems containing \( \Omega_{3c} \) baryons and nucleons. 
The analysis suggests the possible existence of bound \( \Omega_{3c} \) states 
within wide energy ranges. Finally, we must notice that the results have a hypothetical character due to the lack of appropriate experimental data.
 
%=================================================
 \subsection{$\Omega \alpha$ and $\Omega_{3c} \alpha$ potentials}
 
The P$_1$ \( \Omega N \) potential (\ref{HALInteraction}) supports $ \Omega \alpha$ bound
state with a binding energy of approximately 22 MeV and is parameterized in Ref. \cite{Etminan2020,APS25} 
in the form of the Woods-Saxon type potential suggested by Dover and Gal \cite{Gal83}
$
V_{\Omega \alpha}(r)=V_{0}\left[ 1+\exp \left( \frac{r-R}{c}\right) \right]
^{-1},
$
with the strength $V_{0}=-61$ MeV, the surface diffuseness $c$=0.47, and $%
R=1.1A^{1/3}$, where $A$ the mass number of the nuclear core. %In our case $%
%A=4$ and, therefore, $R=1.75.$ %Notice that as there are not sufficient experimental data for the $^5_{\Omega}$He bound state, therefore, the depth of the potential is uncertain. 
Notice that due to the lack of sufficient experimental data for the $^5_{\Omega}$He bound state, the depth of the potential is uncertain \cite{Etminan2020}. 

In contrast to Refs.~\cite{Etminan2020,APS25}, we employ a two-range Gaussian form to parametrize the $\Omega\alpha$ folding potential, as it provides a better approximation in the asymptotic region of the folding potential.
\begin{table}[!ht]
\caption{Low-energy parameters of the $Y\alpha$ potentials: scattering length $a_{Y\alpha}$, effective radius $r_{Y\alpha}$, and ground state energy $E_2$ for $Y = \Omega, \, \Omega_{3c}, \, \Lambda, \, \Xi$. The $\Omega\alpha$ potential is derived from a folding procedure with $\Omega N$ ($\Omega_{3c} N$) potential $\widetilde{V}_{\Omega N}$ ($\widetilde{V}_{\Omega_{3c}N}$) and with an $rms$ radius of 1.70~fm for the $\alpha$-particle matter density. }
\label{t00}
\begin{tabular}{cccccl} \hline\noalign{\smallskip}
$Y \alpha$ potential  &$YN$ potential&Nucleus& $a_{Y \alpha}$ (fm) & $r_{Y\alpha} $ (fm)& $E_2$ (MeV) \\
\hline\noalign{\smallskip}
$\Omega \alpha $ & $\widetilde{V}_{\Omega N}$ (\ref{P1t})  &  $^5_\Omega$He      &  2.73 & 1.05  &-6.16  \\
$\Omega_{3c}\alpha $& $\widetilde{V}_{\Omega_{3c}N}$ (\ref{Vccc}), $c_1$=1.0, $c_2$=1.00  &   $^5_{\Omega_{3c}}$He       &  76.9 & 6.28  & -9.90 \\
$\Omega_{3c}\alpha $& $\widetilde{V}_{\Omega_{3c}N}$ (\ref{Vccc}), $c_1$=1.3, $c_2$=1.35  &     & 2.71 & 1.07 & -3.53 \\
\hline \noalign{\smallskip} 
 $\Lambda  \alpha$ \cite{Isle}&-& $^5_\Lambda$He&4.2 &1.9& -3.10 \cite{FSV09}\\
$\Xi  \alpha$ \cite{GVV16,FSV08}& - &$^5_\Xi$He  & 4.6 & 2.2 & -2.09 \cite{FSV17} \\
\noalign{\smallskip}\hline
\end{tabular}
\end{table}

Table~\ref{t00} lists the low-energy properties of $\Omega\alpha$ interactions, 
where $Y = \Omega$ or $\Omega_{3c}$. Shown are the scattering length $a_{\Omega\alpha}$, 
effective range $r_{\Omega\alpha}$, and two-body ground-state energy $E_2$. 
The $\Omega\alpha$ and $\Omega_{3c}\alpha$ potentials are obtained by a Gaussian fit 
in the asymptotic region ($r \ge 2$~fm) to the folding potentials derived from 
the modified $\Omega N$~(\ref{P1t}) and $\Omega_{3c}N$~(\ref{Vccc}) interactions, 
assuming an $\alpha$-particle matter radius of 1.70~fm.
     
For the $\Omega\alpha$ system, the resulting $^5_\Omega$He bound state has a scattering length of 2.73~fm, an effective range of 1.05 fm, and a ground state energy of $-6.16$~MeV. The $\Omega_{3c}\alpha$ interaction is evaluated for two parameter sets of the modified $\Omega N$ potential (\ref{P1t}): for $c_1$=1.0 and $c_2$=1.0, a deeply bound state is observed ($E_2=-9.90$~MeV) with a very large scattering length (76.9 fm), indicating strong attraction. When the parameters are modified to $c_1$=1.3 and $c_2$=1.35, the interaction becomes shallower, leading to $a_{\Omega\alpha}$=2.71~fm, $r_{\Omega\alpha}$=1.07~fm, and $E_2$=-3.53~MeV.
That is comparing to known results for the $\Lambda\alpha$ and $\Xi\alpha$ systems from Refs.~\cite{Isle,FSV09,GVV16,FSV08,FSV17}, which exhibit binding energies of 3.10~MeV and 2.09~MeV, respectively.  

The $\Omega\alpha$ and  $\Omega_{3c}\alpha$ ($c_1=1.3$, $c_2=1.35$) potentials are parameterized as a sum of two Gaussian terms:  
\begin{equation}
V_{\Omega\alpha}(r)=-1.288\exp(-(r/1.153)^2)-23.998\exp(-(r/2.624)^2),
\label{Omega_alpha}
\end{equation}
\begin{equation}
V_{\Omega_{3c}\alpha}(r)=-0.0206\exp(-(r/1.544)^2)-14.219\exp(-(r/2.527)^2), \quad
\label{Omega3c_alpha}
\end{equation}
where $V_{\Omega\alpha}(r)$ ($V_{\Omega_{3c}\alpha}(r)$) is in MeV and $r$ in fm.
The results  shows that the potential (\ref{Omega_alpha}) with parameter set (\ref{P1t}) of the $\Omega N$ potential  leads to the strongly bound ${^{5}_{\Omega}}$He nucleus, while ${^{5}_{\Lambda}}$He and ${^{5}_{\Xi}}$He are significantly less bound. At the same time, the binding energy of ${^{5}_{\Omega_{3c}}}$He can be compareble with the energy of the hypernuclei.

Overall, results in Table \ref{t00} illustrates that $\Omega$ and $\Omega_{3c}$ baryons can form bound states with the $\alpha$ particle under our potential assumptions, in some cases with stronger binding than conventional hyperons from
the baryon octet. 

\subsection{$\Omega_{3x}$-containing $\alpha$-cluster systems}
%\subsection{$\alpha$-cluster system with $\Omega$ baryons}
Let us first consider calculations for ${^{6}_{\Omega\Omega}}$He hypernucleus  formed by the $\Omega\Omega\alpha$ system. Results of different characteristis for this system are presented in Tables \ref{t22}-\ref{t44}. For the comparison, there are also presented the corresponding characteristics for ${^{6}_{\Lambda\Lambda}}$He and  ${^{6}_{\Xi\Xi}}$He hypernuclei. The corresponding calculations for  ${^{6}_{\Lambda\Lambda}}$He and  ${^{6}_{\Xi\Xi}}$He hypernuclei were performed in Refs. \cite{FGS04,EH22,FSV17} using the cluster model for $\Lambda\Lambda\alpha$ and $\Xi\Xi\alpha$ systems in the framework of the Faddeev equations in configuration space. 
%=============================================

\begin{table}[!ht]
\caption{Ground state energy of several baryon systems including a single $\alpha$-cluster. The calculated (experimantal) energy of ground state $E_3$ ($E_3^{exp.}$). The notation is the same as in Table \ref{t00}. 
}
\label{t44}
\begin{tabular}{lccc} \hline\noalign{\smallskip}
Model& Nucleus&  $E_3$ (MeV) &  $E_3^{exp.}$ (MeV) \\
\hline\noalign{\smallskip}
$\Omega \Omega \alpha $, $\Omega \alpha $ potential (\ref{Omega_alpha})&$_{\Omega\Omega}^6$He &-21.93 & --\\
$\Omega_{3c} \Omega_{3c}\alpha $, $\widetilde{V}_{\Omega_{3c}N}$  (\ref{Vccc}), $c_1$=1.0, $c_2$=1.0, &$_{\Omega_{3c}\Omega_{3c} }^6$He &-34.95 & --\\
 \hline \noalign{\smallskip}
$\Lambda \Lambda \alpha$& $^6_{\Lambda\Lambda}$He&  -6.903 \cite{FGS04}   & -7.25 \cite{Exp_lambda}\\
&&  -7.468\cite{EH22}  & \\ 
$\Xi \Xi \alpha$ &$^6_{\Xi\Xi}$He  &-7.635  \cite{FSV17}& -- \\ \hline \noalign{\smallskip} 
%$ N N \alpha$ (+3BF) &$^6$He& -0.973\cite{FSV14} & -0.973\\   
%\noalign{\smallskip}\hline
\end{tabular}
\end{table}
Table \ref{t44} is listing  the calculated ground state energy $E_3$ of various baryon systems.  The experimental values  are provided where available. 
In Table \ref{t44} Results for the ground state energy of the ${^{6}_{\Omega\Omega}}$He and $^{6}_{\Omega_{3c}\Omega_{3c}}$He nuclei, obtained within a three-body $\Omega\Omega\alpha$ and $\Omega_{3c}\Omega_{3c}\alpha$ cluster model, are presented along with the energies of ${^{6}_{\Lambda\Lambda}}$He, ${^{6}_{\Xi\Xi}}$He.  The binding energy of $^{6}_{\Omega_{3x}}$He is almost three-five times bigger than the corresponding energy values for ${^{6}_{\Lambda\Lambda}}$He, ${^{6}_{\Xi\Xi}}$He.  
    
The bound state energies $E_3$ are calculated within the $\Omega \alpha $ potential (\ref{Omega_alpha}) and  $\Omega_{3c} \alpha $ potential without scalling $\Omega_{3c} N $ interaction. 
For comparison, we also include results for the double $\Lambda$ hypernucleus ($_{\Lambda\Lambda}^6\mathrm{He}$), where both theoretical and experimental $E_3$ values are available. 
In the case of the double $\Xi$ hypernucleus ($_{\Xi\Xi}^6\mathrm{He}$), only a theoretical prediction is reported. 
 The properties of the $\Omega \Omega\alpha$ system depend on theoretical models, such as lattice QCD predictions and phenomenological folding potentials. Given that the $\Omega\Omega$ interaction is sufficiently attractive, a bound state of $\Omega\Omega\alpha$ may exist. A unique aspect of the $\Omega\Omega\alpha$ system is that the $\Omega\alpha$ interaction is also attractive and forms a bound $\Omega\alpha$ pair. This differs from the $\alpha N$ interaction in the $\alpha NN$ system due to the Pauli exclusion principle. Consequently, one might expect a larger binding energy in the $\Omega \Omega\alpha$ system.

\begin{table}[!ht]
\caption{
Ground state energy of $Y\alpha\alpha$ system, $Y=\Omega$, $\Omega_{3c}$,  $\Lambda$, $\Xi$, calculated with the Ali-Bodmer $\alpha\alpha$ potential (AB(a)–AB(d) mixing) and orbital configurations $(l_{Y\alpha}, l_{\alpha\alpha}) = {(1,2,3), (0,2,4)}$.  The notation is the same as in Table \ref{t00}.
}
\label{t33}
\begin{tabular}{lccc} \hline\noalign{\smallskip}
Model & Nucleus&  $E_3$ (MeV) &  $E_3^{exp.}$ (MeV) \\
\hline\noalign{\smallskip}
$\Omega \alpha \alpha  $, $\Omega \alpha $ potential (\ref{Omega_alpha})&$^9_\Omega$Be    & -12.9& --\\ 
$\Omega_{3c}\alpha \alpha$, $\widetilde{V}_{\Omega_{3c}N}$  (\ref{Vccc}), $c_1=1.0$, $c_2=1.0$&$^9_{\Omega_{3c}}$Be   & -18.0& --\\ 
$\Omega_{3c} \alpha \alpha$,  $\Omega_{3c} \alpha $ potential (\ref{Omega3c_alpha}) && -7.39& --\\ 
 \hline \noalign{\smallskip} 
 $\Lambda \alpha \alpha$, Gibson $\Lambda \alpha$ potential& $^9_\Lambda$Be&  -6.71\cite{FGS04} & -6.62\cite{Exp_Laa}\\ 
$\Xi \alpha \alpha$, Isle  $\Xi  \alpha$ potential \cite{FSV17}&$^9_\Xi$Be  &-6.04& --\\ \hline \noalign{\smallskip} 
\end{tabular}
\end{table}
Table \ref{t33} presents the calculated ground state energies of various light hypernuclear systems of the form $Y\alpha\alpha$, where $Y$ denotes the particles ($\Omega$, $\Omega_{3c}$, $\Lambda$, $\Xi$).  The orbital angular momentum configurations considered in the calculations include partial waves up to $(l_{Y\alpha}, l_{\alpha\alpha}) = ({1,2,3}, {0,2,4})$, ensuring a comprehensive treatment of the low-lying states  in subsystwms.
For the $\Omega\alpha$ interaction, a folding potential is constructed and regularized by fitting the tail of the potential with Gaussian functions, as discussed above. This potential is based on the long-range behavior of lattice QCD-derived $\Omega N$ interactions, with the short-range core ($r < 0.4$ fm) suppressed to avoid unphysical deep binding. The same regularization strategy is extended to the hypothetical $\Omega_{3c}$$\alpha$ systems. Two parameter sets for the Gaussian fits (labeled by coefficients $c_1$ and $c_2$) are used to explore the sensitivity of the binding energy to the $\Omega_{3c}\alpha$ interaction strength.
The table compares these newly predicted bound states with known hypernuclear systems such as $\Lambda$$\alpha$$\alpha$ and $\Xi$$\alpha$$\alpha$, modeled using phenomenological potentials (Gibson for $\Lambda\alpha$ and Isle for $\Xi\alpha$). For reference, the ordinary nucleon $\alpha$ $\alpha$ system ($^9$Be) is also included, highlighting the relative binding strength induced by different hyperons.
We can constatate that \\
  (1)  the $\Omega \alpha \alpha$ system shows strong bound state ($-12.9$ MeV). That, however, does not agree with earlier predictions using lattice QCD-based $\Omega N$ interactions that generate deeply bound states about $-34$ MeV \cite{APS25}.
  This fact can be explaned by the proposed regularization of the interaction.\\
  (2)  The $\Omega_{3c}\alpha\alpha$ system exhibits even stronger binding under moderate $\Omega_{3c}N$  parameters ($c_1=c_2=1.0$), while tuning the interaction to $c_1=1.3$, $c_2=1.35$ results in significantly less binding, illustrating the sensitivity of the cluster to the assumed parameter for the $\Omega_{3c}N$ potential.\\
The $\Lambda \alpha\alpha$ system matches well with experimental data, validating the model setup. 
    The $\Lambda$ and $\Xi$ cases serve as comparative benchmarks, showing relatively weaker binding.
Overall, this table illustrates the predictive power of the regularized $\Omega N$ and $\Omega_{3c}N$ potentials in generating bound cluster states  and supports the broader conclusion that such systems are viable candidates for experimental searches.

\subsection{$\Omega\Omega$ potential at short distances}

It was assumed that when the distance between the centers of mass of two $\Omega$s is smaller than twice the radius of the $\Omega$ baryon, the description of the system in terms of independent particles interacting via pairwise potentials becomes invalid. Therefore, it is essential to identify the region of inter-particle distances where lattice QCD-based potentials can be applied with confidence. Taking the radius of the $\Omega$ baryon to be approximately $0.6$~fm, we define a characteristic threshold distance $r_t = 1.2$~fm. 

The $rms$ radii of some baryons can be found in Refs. \cite{Alexandrou,Yu2023}
From lattice QCD calculations by Alexandrou \textit{et al.} \cite{Alexandrou}, 
the electric charge $rms$ radius of the $\Omega$ baryon  is given as
$
\sqrt{\,|\langle r^{2}_{E0} \rangle|\,} \approx 0.595 \pm 0.008~\mathrm{fm}.
$
Thus, the $rms$ charge radius of the $\Omega^{-}$ baryon is approximately 
$0.60~\mathrm{fm}$.

Guo-Liang Yu \textit{et al.} \cite{Yu2023} employed a relativized quark model 
to compute $rms$ radii (among other properties) of 
singly heavy baryons such as $\Omega_{Q}$.
Although precise numbers are not explicitly quoted, these model-based 
$rms$ radii typically fall in the range of $\sim 0.5$--$0.7~\mathrm{fm}$, 
depending on the baryon mass and the spatial configuration of the 
constituent quarks.

 In the Faddeev formalism, when particles are treated as point-like, structureless objects, it is necessary to distinguish between two interaction regions. The asymptotic region is defined as $r \ge r_t$, where the interaction corresponds directly to the $\Omega\Omega$ potential as derived from lattice QCD. However, for smaller distances, the interaction must be understood as arising from the underlying quark dynamics.
The  two-body interaction of two point-like particle becomes to be unrealistic. This part of the $\Omega\Omega$ potential can be considered as an effective pair potential.
 The repulsive short-range component of the $\Omega\Omega$ interaction~(\ref{OmegaOmega}) is usually attributed to Pauli repulsion. 

 Let us draw an analogy with the nucleon-nucleon interaction. 
It is well known that realistic phenomenological $NN$ potentials can successfully describe two-body scattering data \cite{RNHF}. 
However, these same potentials fail to reproduce three-body data with the same level of accuracy. 
This issue is typically resolved by introducing three-body forces, which effectively account for short-range corrections to interaction
of three nucleons.
In the same way, the results obtained for the present three-body system like  $\Omega NN$ or  $\Omega \Omega N$ are likely to change once experimental data become available.

A further example is provided by the short-range contact component of the chiral EFT 
$NN$ interaction, which at leading (LO) and next-to-leading order (NLO) is represented 
by contact operators formulated in the standard momentum-space operator basis of modern 
chiral potentials. In coordinate space, these contact terms are typically regularized 
by introducing a smeared $\delta$ function:  
\begin{equation}
\delta_\Lambda(\mathbf{r}) = \left(\frac{\Lambda}{\sqrt{\pi}}\right)^3 e^{-\Lambda^2 r^2},
\quad
V_{\text{LO}}(\mathbf{r}) = \big(C_S + C_T \,\boldsymbol{\sigma}_1\!\cdot\!\boldsymbol{\sigma}_2\big)\, 
\delta_\Lambda(\mathbf{r}),
\label{cont}
\end{equation}
which effectively suppresses the unphysical overlap of nucleon wave functions at short 
distances. In this formula, the constants $C_S$ and $C_T$ are the low-energy constants 
that parametrize the short-range part of the nucleon - nucleon  interaction 
at LO in chiral EFT.  The $\Lambda$ is the cutoff scale of the regulator, introduced to make the $\delta$ 
function finite-ranged and to separate low-energy physics (treated explicitly) 
from unresolved short-distance physics (absorbed into the constants $C_S, C_T$).
In practical nuclear EFT, typical values are 
$\Lambda \sim 450 \text{--} 600~\text{MeV}$, 
corresponding to coordinate-space regulator radii 
$R \sim 0.3 \text{--} 0.5~\text{fm}$, 
consistent with the approximate relation $R \sim 1/\Lambda$.
\begin{figure}[t]
\begin{center}
\includegraphics[width=25pc]{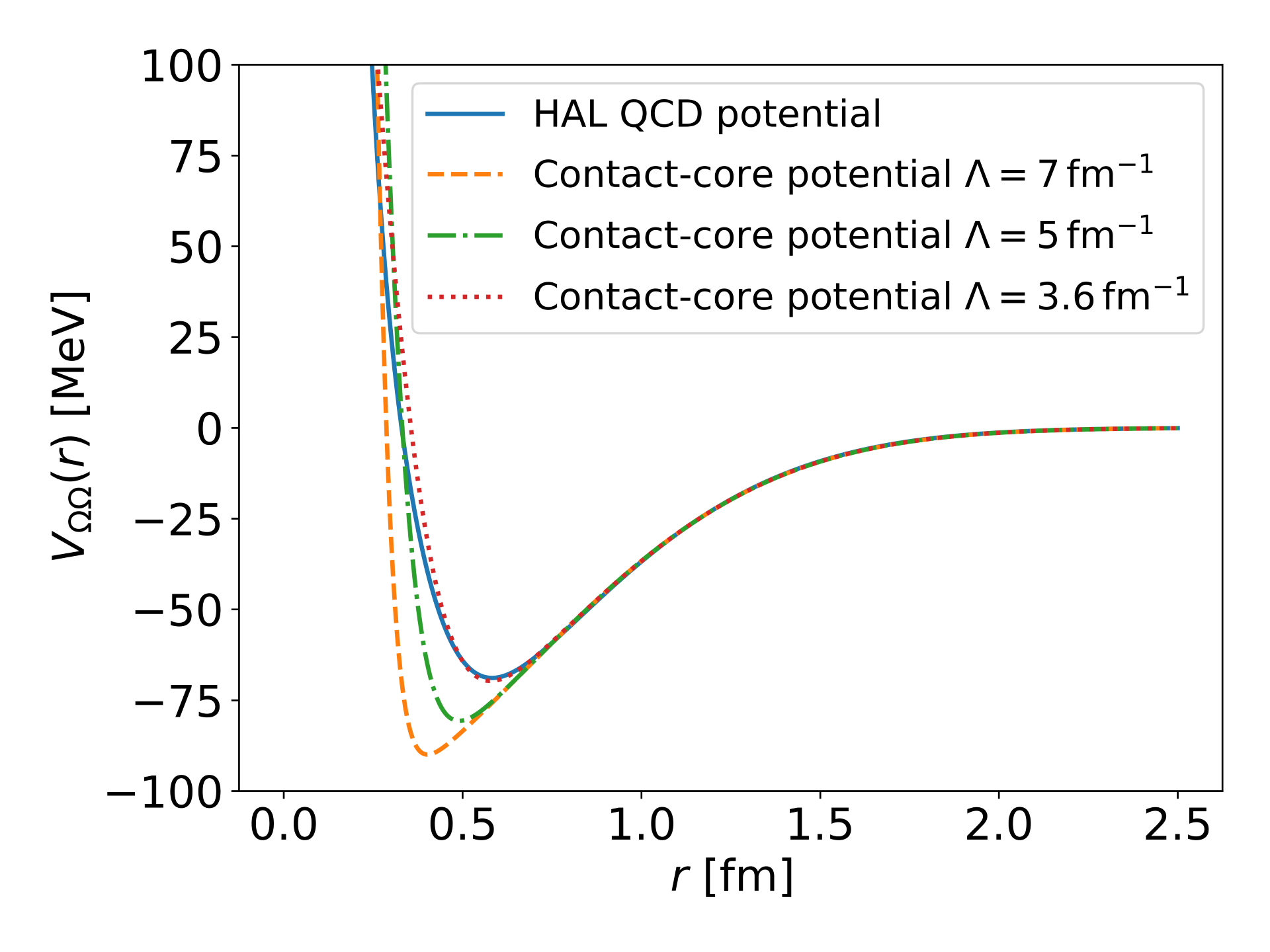}
\end{center}
\caption{\label{fig:7} The $\Omega\Omega$ potential (solid curve) compared with contact-like simulation potentials: $\Lambda = 7.0$ fm$^{-1}$ (dashed curve), $\Lambda = 5.0$ fm$^{-1}$ (dot-dashed curve), and $\Lambda = 3.6$ fm$^{-1}$ (dotted curve).
}
\end{figure}

According to Eq.~(\ref{VV}), we attempt to decompose the $\Omega\Omega$ potential 
into attractive and repulsive parts. The attractive component is fitted with a Gaussian 
function, while the repulsive component is represented by a contact potential of the form 
Eq.~(\ref{cont}) \cite{AvB},  
$
V(r) = C_\Lambda \, e^{-\Lambda^2 r^2},
$ 
where $C_\Lambda$ is a parameter that depends on $\Lambda$, and $\Lambda$ is expressed in fm$^{-1}$.  

The new potential reproduces the same binding energy as the original potential by 
appropriately adjusting the parameters $C_\Lambda$ and $\Lambda$. A comparison of the original 
and parametrized potentials is shown in Fig.~\ref{fig:7}. The regulator radii $R$ were 
varied as $0.14$~fm, $0.20$~fm, and $0.27$~fm, which correspond to cutoff values of 
$\Lambda = 1142$~MeV, $585$~MeV, and $301$~MeV, respectively.  
It is not surprising that these simulated potentials reproduce the three-body bound state 
energy obtained with the original $\Omega\Omega$ potential (see Table~\ref{t111}) with 
comparable accuracy. In these calculations, the $\Omega N$ potential P$_1$ was employed. 
Thus, we obtain a set of repulsive cores for the $\Omega\Omega$ interaction that yield 
the same two- and three-body bound state energies. While the obtained potentials differ at 
short distances, they exhibit the same asymptotic behavior at medium and large distances.

One can conclude that the $\Omega\Omega$ potential is not well defined (or cannot be properly described by a pairwise potential) at short distances, due to the fundamental difference between the particle-based picture at asymptotic scales and the underlying quark dynamics at short distances.

Finally, the explicit form of the $\Omega\Omega$ interaction employed in the calculations is given by
the contact-core simulating potentials
\[
V_{\Omega\Omega}(\Lambda,C_\Lambda,r)= -109.93 \, e^{-(r / 0.955)^2} 
+ \left(\frac{\Lambda}{\sqrt{\pi}}\right)^3 e^{-\Lambda^2 r^2} \, C_\Lambda ,
\]
with the following parameter sets:
$
\Lambda = 7~\text{fm}^{-1},  C_\Lambda= 93~\text{MeV fm}^3;\quad
$
$
\Lambda = 5~\text{fm}^{-1},  C_\Lambda= 68~\text{MeV fm}^3;\quad
$
$
\Lambda= 3.6~\text{fm}^{-1}, C_\Lambda = 59~\text{MeV fm}^3.
$

\section{Concluding remarks}
\label{sec:6}

The models employed for the $\Omega N$ and $\Omega_{3c}N$ interactions, though derived from or inspired by lattice QCD, may be overly attractive when applied to few-body cluster systems such as $\Omega \alpha$ and $\Omega_{3c}\alpha$. This excessive attraction can lead to unrealistically deep binding energies that may not be confirmed by future experimental data. A refined treatment of these interactions is therefore necessary, particularly for studies of few-body dynamics and the possible appearance of overbound states.  

To address the overly attractive core of the HAL QCD $\Omega N$ potential, we applied a mitigation procedure in which its asymptotic region ($r \geq 1.2$~fm) was fitted with a two-range Gaussian parametrization. The resulting interaction preserves the low-energy properties of the original potential. The renormalized form was tested by calculating the ground-state energies of the three-body systems $\Omega NN$ and $\Omega \Omega N$, yielding results consistent with those obtained using the unmodified HAL QCD potential.  

Furthermore, by applying a folding procedure to the renormalized $\Omega N$ interaction, we derived effective $\Omega \alpha$ and $\Omega_{3c}\alpha$ potentials for use in few-body cluster calculations. These folding potentials were then fitted in the asymptotic region ($r \geq 2.0$~fm), defined by the $rms$ radius of the $\alpha$ particle, using Gaussian functions.  

Our results indicate that the large binding energies predicted for the hypothetical clusters $^9_\Omega$Be and $^6_{\Omega\Omega}$H arise primarily from the strong $\Omega \alpha$ attraction, with the $\Omega\alpha$ subsystem itself forming a bound state of about $-20$~MeV. It should be noted, however, that folding procedures generally overbind cluster systems. A more precise determination of the effective potential therefore requires experimental input on binding energies in order to correct for the limitations of the folding approximation.  

Similarly, the predicted binding of $^6_{\Omega\Omega}$He originates from the strong $\Omega\Omega$ attraction, which supports a dibaryon bound state.  

For $\Omega_{3c}$-containing systems, we introduced an $\Omega_{3c}N$ interaction derived from the $\Omega N$ potential, guided by the relation between the $\Omega\Omega$ and $\Omega_{3c}\Omega_{3c}$ interactions. Several parametrizations of this interaction predict bound states in the corresponding systems.  

Finally, the $\Omega\Omega$ interaction, Eq.~(\ref{OmegaOmega}), was analyzed within the proposed framework. Its short-range repulsion, likely associated with Pauli effects, dominates at $r \lesssim 0.4$~fm, while the intermediate- and long-range attraction generates a two-body bound state of about $1.41$~MeV and a three-body state at $6.0$~MeV in $\Omega\Omega N$. The repulsive core can be effectively modeled by contact-like interactions with adjustable cutoff and strength parameters, which reproduce the same binding energies as the original lattice-derived potential. This highlights a fundamental tension between the particle-based description valid at large distances and the underlying quark dynamics that govern short-range physics.  

\section*{Acknowledgments}

This work is supported by the City University of New York, PSC CUNY Research Award No 68511-00-56 and Department of Energy$/$National Nuclear Security
Administration Award No NA0003979.

\end{document}